\documentstyle{article}
\title{Green functions and dimensional reduction of quantum fields on product manifolds }
\author{ Z. Haba\\Institute of Theoretical Physics, University of Wroclaw,
\\50-204 Wroclaw, Plac Maxa Borna 9, Poland\\e-mail:zhab@ift.uni.wroc.pl}
\date{ }
\begin{document}
\maketitle
\begin{abstract}
We discuss Euclidean  Green functions on product manifolds ${\cal
P}={\cal N}\times {\cal M}$. We show that if ${\cal M}$ is compact
and ${\cal N}$ is not compact then the Euclidean field on ${\cal P}$ can be approximated by its
zero mode which is a Euclidean field on ${\cal N}$. We estimate
the remainder of this approximation. We show that for large
distances on ${\cal N}$ the remainder is small. If ${\cal
P}=R^{D-1}\times S^{\beta}$, where $S^{\beta}$ is a circle of
radius $\beta$, then the result reduces to the well-known
approximation of the $D$ dimensional finite temperature quantum
field theory by $D-1$ dimensional one in the high temperature
limit. Analytic continuation of Euclidean fields is discussed
briefly.\end{abstract}

\section{Introduction}The aim of the Kaluza-Klein  program \cite{KK}
is a unification of interactions by means of an extension of the
number of dimensions $D$. Then, $D-4$ dimensions are supposed to
be unobservable. In this paper we ask the question whether a
higher dimensional quantum field theory can be approximated  by a
lower dimensional one (not necessarily from the point of view of
the Kaluza-Klein program). We have shown in our earlier paper
\cite{haba} (see also \cite{padma}) that such an approximation
applies near the bifurcate Killing horizon when $D$ dimensional
quantum field theory can be approximated by a two-dimensional one.
We are interested in the dimensional reduction from the point of
view of Green functions,i.e., correlation functions of quantum
fields. We examine the question whether the compact dimensions are
negligible. In terms of the Green functions this means that Green
functions in higher dimensions should be approximated by the ones
in lower dimensions. Such a property cannot be true at arbitrarily
small distances (except of some brane-type models
\cite{warpbrane}\cite{dvali}) because the singularity of the Green
function depends on the dimension. So, the approximation can make
sense only above a certain length scale. If the manifold ${\cal
P}$ is the product manifold ${\cal P}={\cal N}\times {\cal M}$
then we expect that the Green functions on ${\cal P}$ can be
approximated by the ones on ${\cal N}$ if the distances on ${\cal
N}$ are big in comparison to the size of ${\cal M}$. In the
conventional approach to Kaluza-Klein theories \cite{KK}, the
Fourier expansion of functions on the compact manifold ${\cal M}$
leads to large masses which make (by a formal argument)
propagators of the massive Kaluza-Klein particles negligible,
realizing in this way the dimensional reduction. We are interested
to see how this decoupling works in the configuration space.

In order to study the problem in a rigorous way we work in the
Euclidean (Riemannian instead of pseudoRiemannian) formulation of
quantum field theory. Although the Euclidean approach to fields
on a manifold is not as complete as in the flat space there are
already some crucial results concerning the analytic continuation
and construction of quantum fields \cite{dimock}\cite{jaffe}. The
quantum fields are determined by the Green functions. We discuss
in this paper only the two-point function which is sufficient for
a construction of free fields. The analytic continuation of
interacting Euclidean fields can be performed if ${\cal P}$ and
the interaction have an additional reflection symmetry
\cite{dimock}\cite{jaffe}.

   In sec.2 we define the Green functions. The Green functions are
   expanded in eigenfunctions in sec.3. We distinguish the
   contribution of the zero mode which determines the dimensional
   reduction. In sec.4 we discuss a special case of two
   dimensional manifold ${\cal N}$.We estimate the correction to the
   lower dimensional approximation in sec.5. In sec.6 we construct quantum fields from the Green
   functions. We discuss a possible extension of the results in
   sec.7.

\section{Warped metric on a product manifold}
We consider a manifold in the form of a product ${\cal P}={\cal
N}\times {\cal M}$ where ${\cal N}$ has $D-d$ dimensions and
${\cal M}$ is a $d$ dimensional manifold . We assume that a metric
on ${\cal P}$ can be expressed in the warped form \cite{warp}
\begin{equation}
ds^{2}=\sigma_{AB}dX^{A}dX^{B}=g_{ab}(x)dx^{a}dx^{b}+v^{2}(x)h_{jk}(y)dy^{j}dy^{k}
\end{equation}
where the coordinates  on ${\cal P}$ are denoted by the capital
$X=(x,y)$, the ones on ${\cal N}$ by $x$ and the coordinates on
${\cal M}$ are denoted by $y$. If $v=1$ then the metric on the
product manifold is just a product of the metrics.

Let
\begin{equation}
\triangle_{P}=\frac{1}{\sqrt{\sigma}}\partial_{A}\sigma^{AB}\sqrt{\sigma}\partial_{B}
\end{equation}
be the Laplace-Beltrami operator on ${\cal P}$. We are interested in
the calculation of the Green functions ($\sigma=\det(\sigma_{AB})$)
\begin{equation}
(-\triangle_{P} +m^{2}){\cal
G}^{m}=\frac{1}{\sqrt{\sigma}}\delta
\end{equation}
In the metric (1) eq.(3) reads
\begin{equation}
\Big(-v^{d-2}\sqrt{g}(\triangle_{M}-m^{2})-
\partial_{a}g^{ab}v^{d}\sqrt{g}\partial_{b}\Big){\cal
G}^{m}=h_{M}^{-\frac{1}{2}}\delta(X-X^{\prime})
\end{equation}
A solution of eq.(3) can be expressed by the fundamental solution
 of the diffusion equation
\begin{equation}
\partial_{\tau}P_{\tau}=\frac{1}{2}\triangle_{P}P_{\tau}
\end{equation}
with the initial condition
$P_{0}(X,X^{\prime})=\sigma^{-\frac{1}{2}}\delta(X-X^{\prime})$.
Then
\begin{equation}
{\cal
G}^{m}=\frac{1}{2}\int_{0}^{\infty}d\tau\exp(-\frac{1}{2}m^{2}\tau)
P_{\tau}
\end{equation}If $v=1$ then we may write eq.(4) in the form (here $
h_{M}=\det(h_{jk})$ and $g=\det(g_{ab})$)
\begin{equation}
(-\triangle_{M}+m^{2}-\triangle_{N} ){\cal
G}^{m}=h_{M}^{-\frac{1}{2}}g^{-\frac{1}{2}}\delta( X-
X^{\prime})
\end{equation}
In such a case  from eq.(5) we obtain a  simple formula ( in the sense of a product of
semigroups)
\begin{equation}
P_{\tau}^{P}=P_{\tau}^{N}P_{\tau}^{M} \end{equation}
where the upper index of the heat kernel denotes the manifold of its definition.

 Hence
\begin{equation}
{\cal G}^{m}(X,X^{\prime})=\frac{1}{2}\int_{0}^{\infty}d\tau\exp(-\frac{1}{2}m^{2}\tau)
P_{\tau}^{N}(x,x^{\prime})P_{\tau}^{M}(y,y^{\prime})
\end{equation}
The formula (9) is useful  if we have a reliable approximation for
the heat kernels on ${\cal N}$ and ${\cal M}$. We could conclude
from eq.(9) (using the Schwinger-DeWitt asymptotic expansion) that
if $X$ is close to $X^{\prime}$ then ${\cal G}\simeq
(s_{N}^{2}(x,x^{\prime})
+s_{M}^{2}(y,y^{\prime}))^{-\frac{D}{2}+1}$, where $s_{M}$ denotes
the geodesic distance on ${\cal M}$. We are interested in the
behaviour of the Green functions when  $s_{N}>>s_{M}$. For such a
purpose the formula (9) does not seem  useful. We apply
eigenfunction expansions of the heat kernels in the next section.

 The special case of $D-d=2$ and $v=1$ can be studied in more
 detail.
We choose the isometric coordinates with $g_{ab}=\delta_{ab}a^{2}$.
In such a case eq.(4) reads
\begin{equation} \Big(a^{2}(x)(-\triangle_{M}+m^{2})-
\triangle_{2}\Big){\cal
G}^{m}=h_{M}^{-\frac{1}{2}}\delta(X-X^{\prime})
\end{equation}
where $\triangle_{2}$ is the Laplacian on $R^{2}$.

\section{Eigenfunction expansions}
We assume in this section that ${\cal M}$ is a compact manifold
without a boundary. Then, $-\triangle_{M}$ has a complete discrete
set of orthonormal eigenfunctions \cite{chavel}
\begin{equation} -\triangle_{M}u_{k}=\epsilon_{k}u_{k}
\end{equation}
satisfying the completeness relation
\begin{equation}
\sum_{k}\overline{u}_{k}(y)u_{k}(y^{\prime})=h_{M}^{-\frac{1}{2}}\delta(y-y^{\prime})
\end{equation}
Let us note that $1$ is an eigenfunction (11) with the eigenvalue
$0$ (we use the normalization $\int dy\sqrt{h_{M}}=1$).
Then, (distinguishing the zero mode) we can expand the heat kernel
in eigenfunctions
\begin{equation}
P_{\tau}^{M}(y,y^{\prime})=1+\sum_{k\neq
0}\exp(-\frac{1}{2}\epsilon_{k}\tau)\overline{u}_{k}(y)u_{k}(y^{\prime})
\end{equation}
For the ${\cal N}$ part of ${\cal P}$ we consider the eigenvalue
problem in $L^{2}(dx)$ suggested by eq.(4)
\begin{equation}
{\cal A}_{k}\phi_{E}^{k}=\Big(v^{d-2}\sqrt{g}\omega_{k}^{2}-
\partial_{a}g^{ab}v^{d}\sqrt{g}\partial_{b}\Big)\phi^{k}_{E}=E_{k}\phi_{E}^{k}
\end{equation}where
 \begin{equation}
  \omega_{k}^{2}=\epsilon_{k}+m^{2}
  \end{equation}
The eigenfunctions satisfy
  the  completeness relation
  \begin{displaymath}
\sum_{E}\overline{\phi}_{E}^{k}(x)\phi_{E}^{k}(x^{\prime})
 =\delta(x-x^{\prime})\end{displaymath}
 where the sum must be replaced by an integral if the spectrum of ${\cal A}_{k}$
 in eq.(14) is continuous.

We expand the Green function in eigenfunctions $u_{k}$ of the
Laplace-Beltrami operator $\triangle_{M}$
\begin{equation}\begin{array}{l}
{\cal G}^{m}(X,X^{\prime})\equiv
\sum_{k}{\cal G}^{m}_{k}(X,X^{\prime}) = \sum_{k}g^{m}_{k}(x,x^{\prime})\overline{u}_{k}(y)u_{k}(y^{\prime})\end{array}
\end{equation}
Then, $g_{k}$ is expanded in the eigenfunctions (14)
\begin{equation}
g^{m}_{k}(x,x^{\prime})=\sum_{E}E_{k}^{-1}\overline{\phi}_{E}^{k}(x)\phi_{E}^{k}(x^{\prime})
 \end{equation}
 $g_{k}^{m}$ is a solution of the equation
 \begin{equation}
 {\cal A}_{k}g_{k}^{m}(x,x^{\prime})=\delta(x-x^{\prime})
 \end{equation}
  $1$ is an eigenfunction of $-\triangle_{M}$
with the eigenvalue $0$. If we subtract the zero mode ${\cal G}^{m}_{0}$ (corresponding
to $u_{0}=1$)  from ${\cal
G}^{m}$ then

 \begin{equation}\begin{array}{l}
{\cal G}^{m}(X,X^{\prime})-{\cal
G}^{m}_{0}(X,X^{\prime})=\sum_{E,k>0}E_{k}^{-1}\overline{\phi}_{E}^{k}(x)\phi_{E}^{k}(x^{\prime})\overline{u}_{k}(y
)u_{k}(y)\end{array}
 \end{equation}

with

\begin{equation}\begin{array}{l}{\cal G}^{m}_{0}(X,X^{\prime})=
 G^{m}(x,x^{\prime})
\end{array}\end{equation}
where $G^{m}$ is a solution of the equation
\begin{equation}
\Big(-
\partial_{a}g^{ab}v^{d}g^{\frac{1}{2}}\partial_{b}+m^{2}v^{d-2}g^{\frac{1}{2}}\Big)G^{m}
=\delta(x-x^{\prime})
\end{equation}
We investigate in this paper whether ${\cal G}^{m}$ can be approximated by ${\cal G}_{0}^{m}$,i.e.,
by the Green function $G^{m}$ on ${\cal N}$. Such an approximation cannot be true for small
distances because if the metric tensor is a regular function then the singularity of the
Green function depends on the dimension of the space-time (see the discussion in sec.7).
We expect that the approximation makes sense for large distances in ${\cal N}$.
It can be seen that a decay of eigenfunctions
$\phi_{E}^{k}$ is sufficient for  a disappearance of each term on the rhs of eq.(19)
at large $x$(this is not a necessary condition as we show soon).
The eigenfunctions $\phi_{E}^{k}$ are
localized  if the spectrum of the operators ${\cal A}_{k}$ (14) is discrete.
 We can estimate the decay of eigenfunctions
$\phi_{E}^{k}$ applying the eikonal (WKB) approximation to eq.(14).
This means that we write $\phi^{k}=\exp(-\omega_{k}W)$ assuming that $W$ is
growing uniformly in each direction for large distances. Then, in
the leading order we obtain for large $x$ the equation
\begin{equation}
1=v^{2}g^{ab}\partial_{a}W\partial_{b}W
\end{equation}
We obtain an exponential localization (increasing with the
eigenvalue $\epsilon_{k}$) of $\phi_{E}^{k}$ if $v^{-2}g_{ab}$ is
uniformly growing to infinity for large distances. Let us note that
$v^{-2}g_{ab}$ is the metric on ${\cal N}$ related to the one of
eq.(1) by a conformal transformation. The growth of
$\tilde{g}_{ab}=v^{-2}g_{ab}$ means that the volume element $\int
dx\sqrt{\tilde{g}}$ is infinite. Such a property could be used to
characterize the manifolds ${\cal P}$ whose Green function is
dominated by the zero mode. Our rough arguments need a
confirmation by a mathematical theory of the eigenvalue problems of
second order differential operators (see \cite{agmon}
for some partial results). The sum over eigenvalues
$\epsilon_{k}$ (the rhs of eq.(19) ) will be discussed in sec.5.

In the special case (10) when $v=1$ and $D-d=2$ the eigenvalue
equation (14) reduces to the well-known problem of quantum mechanics
  \begin{equation}
 {\cal A}\phi_{E}^{k}= (-\triangle_{2}+\omega_{k}^{2}a^{2}(x))\phi_{E}^{k}=
  E_{k}\phi_{E}^{k}
  \end{equation}
Here, $\triangle_{2}$ denotes the two-dimensional Laplacian. In this
special case we have simple criteria for the discreteness of the
spectrum of ${\cal A}$. If $a^{2}$ is growing uniformly in all
directions then the spectrum of ${\cal A}$ is discrete and the
eigenfunctions are localized ( see \cite{simon} for a precise
formulation and proofs).The  eikonal approximation
reads\begin{equation} \nabla W\nabla W=a^{2}(x)
\end{equation}
Hence, $\vert \nabla W(x)\vert$ is growing like $a(x)$.
The decay of eigenfunctions derived from eq.(24) is in agreement with
exact results \cite{simon}. If the eigenfunctions $\phi_{E}^{k}$
decay for large $x$ then the volume of ${\cal N}$, equal to $\int dx a^{2}(x)$,
is infinite (${\cal N}$ is not compact).

The  localization of eigenfunctions $\phi_{E}^{k}(x)$
is not necessary   for a decrease of
${\cal G}^{m}-{\cal G}^{m}_{0}$ .  Let us consider the simplest case of a non-compact ${\cal N}=R^{D-d}$
with $v=1$ and $g_{ab}=\delta_{ab}$.
 Then, ${\cal A}_{k}$ has a continuous spectrum, its eigenfunctions are not localized, but
\begin{equation}\begin{array}{l} ({\cal
G}^{m}-{\cal G}^{m}_{0})(X,X^{\prime})\cr
=\sum_{k \neq
0}\int_{0}^{\infty}d\tau\exp(-\frac{1}{2}\omega_{k}^{2}\tau
-\frac{1}{2\tau}(x-x^{\prime})^{2})(2\pi\tau)^{-\frac{D-d}{2}}
\overline{u}_{k}(y)u_{k}(y^{\prime})\cr\equiv
 \sum_{k
\neq 0}g_{k}^{m}(x-x^{\prime}) \overline{u}_{k}(y)u_{k}(y^{\prime})\end{array}\end{equation} where

\begin{equation}\begin{array}{l} {\cal G}_{0}^{m}(x,x^{\prime})=g_{0}^{m}(x,x^{\prime})
=\int_{0}^{\infty}d\tau\exp(-\frac{1}{2}m^{2}\tau
-\frac{1}{2\tau}(x-x^{\prime})^{2})(2\pi\tau)^{-\frac{D-d}{2}}
\end{array}\end{equation}
and \begin{displaymath}g_{k}^{m}(x-x^{\prime})= 2(2\pi)^{\nu-1}\vert
x-x^{\prime}\vert^{\nu}\omega_{k}^{-\nu}K_{\nu}(\omega_{k}\vert
x-x^{\prime}\vert)
\end{displaymath}
with $\nu=-\frac{D-d}{2}+1$, where $K_{\nu}$ is the modified
Bessel function of the third kind \cite{grad}. From the asymptotic
expansion of $K_{\nu}$ it follows that (for any $m^{2}\geq 0$)
each term on the rhs of eq.(25) is decaying exponentially for
large $\vert x-x^{\prime}\vert$ . The sum on the rhs of eq.(26) will be estimated in sec.5.

If  ${\cal N}$ is a compact manifold without a boundary then the spectrum
$\lambda_{n}$ of the Laplace-Beltrami operator $\triangle_{N}$ on ${\cal N}$ is
discrete
\begin{equation}
-\triangle_{N}\psi_{n}=\lambda_{n}\psi_{n}
\end{equation}
In such a case\begin{equation} P_{\tau}^{N}( x,
x^{\prime})=1+\sum_{n\neq
0}\exp(-\frac{1}{2}\lambda_{n}\tau)\overline{\psi}_{n}( x)\psi_{n}(
x^{\prime})
\end{equation}
Hence, if $v=1$ then from eq.(9)
\begin{equation}\begin{array}{l} ({\cal
G}^{m}-{\cal G}^{m}_{0})(X,X^{\prime}) =
\sum_{k>0}g_{k}(x,x^{\prime})\overline{u}_{k}(y)u_{k}(y^{\prime})\end{array}\end{equation}where
\begin{equation} g_{k}^{m}(x,x^{\prime})=\sum_{n }(\lambda_{n}+\epsilon_{k}+m^{2})^{-1}
\overline{\psi}_{n}(x)\psi_{n}(x^{\prime})\end{equation}

and
\begin{equation}\begin{array}{l} {\cal G}^{m}_{0}(x,x^{\prime})
=G^{m}(x,x^{\prime})=\sum_{n}(\lambda_{n}+m^{2})^{-1}
\overline{\psi}_{n}(x)\psi_{n}(x^{\prime})\end{array}\end{equation}
is the Green function (3) on ${\cal N}$ (solving eq.(21) for $v=1$).

However, if ${\cal N}$ is compact then there is no reason to neglect the rhs of eq.(29).

\section{A two-dimensional manifold ${\cal N}$ with a Killing vector }
We discuss in this section in more detail the product manifold ${\cal P}={\cal N}\times {\cal M}$
($v=1$) when the two-dimensional manifold ${\cal N}$ has a
symmetry generated by a Killing vector $K$.
 In an adapted system of coordinates such that $K=\partial_{1}$
 the metric
can be written in the form
\begin{equation}
ds^{2}=dx_{0}^{2}+a_{1}^{2}(x_{0})dx_{1}^{2}
+\sum_{jk}h_{jk}(y)dy^{j}dy^{k}
\end{equation}
In such a case the equation for the Green function reads
\begin{equation}
-(\partial_{0}a_{1}\partial_{0}+a_{1}^{-1}\partial_{1}^{2}
+a_{1}\triangle_{M}-m^{2}a_{1}){\cal G}^{m}=
h_{M}^{-\frac{1}{2}}\delta
\end{equation}
 We can write the metric in an equivalent form. Let
\begin{equation}
 \hat{x}_{0}=\int dx_{0}(a_{1}(x_{0}))^{-1}
 \end{equation}
Then,
\begin{equation}
ds^{2}=a_{1}^{2}(\hat{x}_{0})\Big(d\hat{x}_{0}^{2}
+dx_{1}^{2}\Big)+ds_{M}^{2}
 \end{equation}
 where $ds_{M}^{2}$ is the metric on ${\cal M}$.
In the new coordinates
\begin{equation}
(-\hat{\partial}_{0}^{2}-\partial_{1}^{2}-
a_{1}^{2}\triangle_{M}+m^{2}a_{1}^{2}){\cal
G}^{m}=h_{M}^{-\frac{1}{2}}\hat{\delta}
\end{equation}
where $\hat{\delta}=\delta(\hat{x}-\hat{x}^{\prime})$ depends on
$\hat{x}$ variables.

As an example, let  $a_{1}(x_{0})= x_{0}$
 then
\begin{equation} \hat{x}_{0}=\ln(x_{0})
\end{equation}
Hence, eq.(36) takes the form
\begin{equation}
(-\hat{\partial}_{0}^{2}-\partial_{1}^{2}-
\exp(2\hat{x}_{0})(\triangle_{M}-m^{2})){\cal G}^{m}=
h_{M}^{-\frac{1}{2}}\hat{\delta}
\end{equation}
  In spite of $a$ vanishing at zero (in the original $x_{0}$ coordinate) it can be checked by
means of a calculation of the curvature tensor $R$ that $R$ is a
continuous function (if ${\cal M}=R^{n}$ then the formula for the
curvature in Bianchi type space-times derived in
\cite{ful}\cite{davis} gives $R=0$). We have discussed the model
(38) in detail in \cite{haba}. It has been shown that the model
serves as an approximation to the Green function on a space-time
with the bifurcate Killing horizon. If ${\cal M}=R^{d}$
then eq.(38) defines the Green function on the Euclidean version of the Rindler space.

   An interesting class of models results from a choice
of a metric which has a power-like singularity at $x_{0}$ when approaching $x_{0}=0$.
We may choose $a_{1}(x_{0})^{2}=\alpha \vert x_{0}\vert^{2\gamma}$
(the curvature tends to infinity at the singularity if $\gamma\neq 1$). Then,
in the coordinates (34) we have
\begin{equation}
a_{1}(x_{0})^{2}=\vert \hat{x}_{0}\vert^{\frac{2\gamma}{1-\gamma}}
\end{equation}
in  eqs.(35)-(36).

 We apply the eigenfunction expansion (23) of sec.3 to the case  when $a_{1}$ depends
only on $x_{0}$. We consider eq.(36) (we omit the hat over $x_{0}$).
We expand the Green function in a complete set of orthonormal
eigenfunctions of the one-dimensional quantum mechanical problem
\begin{equation} (-\partial_{0}^{2}+\omega_{k}^{2}a_{1}(
x_{0})^{2})\phi_{n}^{k}=\lambda_{n}(k)^{2}\phi_{n}^{k}
\end{equation}
where $\omega_{k}$ is defined in eq.(15) and
\begin{equation}
\sum_{n}\overline{\phi}_{n}^{k}(x_{0})\phi_{n}^{k}(x_{0}^{\prime})=\delta(x_{0}-x_{0}^{\prime})
\end{equation}
Then, a solution of the equation for the Green function (10) has
an expansion
\begin{equation}\begin{array}{l}
{\cal G}^{m}(X,X^{\prime}) \cr = \pi^{-1}
\sum_{k,n}\overline{u}_{k}(y)u_{k}(y^{\prime})\overline{\phi}^{k}_{n}(x_{0})\phi_{n}^{k}(x_{0}^{\prime})
 \lambda_{n}(k)^{-1}\exp(-\lambda_{n}(k)\vert x_{1}-x_{1}^{\prime}\vert)
 \end{array}\end{equation}
 The formula (42) follows from eqs.(16)-(17) if we write
\begin{displaymath}
\phi^{k}_{E}(x_{0},x_{1})=\exp(ip_{1}x_{1})\phi_{n}^{k}(x_{0})
\end{displaymath}
with $E_{k}=p_{1}^{2}+\lambda_{n}(k)^{2}$ and
\begin{displaymath}
\sum_{E}=\int dp_{1}\sum_{n}
\end{displaymath}
Then, the integral over $p_{1}$ in eq.(17) leads to eq.(42).

 The eigenfunctions $\phi^{k}$ are decaying exponentially  if $a_{1}$ is growing at infinity, as can be seen, e.g., from
  the WKB approximation (see \cite{simon} for rigorous results)
 \begin{equation}
 \phi_{n}^{k}(x)\simeq\exp(-\omega_{k}\int dx a_{1}(x))
 \end{equation}
Hence, each term in the expansion (42) is decaying exponentially.
Note that according to eqs.(22),(24) and (43) the terms with larger
eigenvalues $\epsilon_{k}$  are decaying faster then the ones with
the lower eigenvalue.

\section{The correction to the contribution of
the zero mode}We expect that in general (for a non-compact manifold ${\cal N}$)
the difference ${\cal G}^{m}-{\cal G}^{m}_{0}$
is negligible for large distances on ${\cal N}$. First, we must estimate
the Green functions $g_{k}^{m}$ (18)
 of the second order differential operators ${\cal A}_{k}$(depending on $\epsilon_{k}$) on ${\cal N}$
 for large distances.
We write
\begin{equation}
g_{k}^{m}(x,x^{\prime})=\exp(-\omega_{k}W(x,x^{\prime}))
\end{equation}
Assuming that $W$ is growing uniformly in each direction we obtain
 in the leading order for large distances eq.(22) for $W$. We recognize eq.(22) as an equation
for a geodesic distance $s^{v}_{N}$ on the manifold ${\cal N}$
with the metric $v^{-2}g_{ab}$ \cite{dewitt}. Hence,
the geodesic distance $W(x,x^{\prime})=s_{N}^{v}(x,x^{\prime})$ is the solution of eq.(22) which is symmetric
under the exchange of the points and satisfies the boundary condition $W(x,x)=0$.
We insert the approximate solutions $g_{k}^{m}$ (44)
(for some rigorous results on the large distance behaviour of Green functions of second
order differential operators see \cite{agmon}\cite{har}) into the sum (19)
 over eigenvalues and eigenfunctions of $-\triangle_{M}$. We approximate the sum over large eigenvalues
 by an integral (Weyl approximation) assuming $\vert u_{k}(y)\vert\leq C$. Then
\begin{equation}\begin{array}{l}
\vert({\cal G}^{m}-{\cal G}^{m}_{0}(X,X^{\prime})\vert\cr
\leq
A_{1}\sum_{k<\Lambda}
  \exp (-\omega_{k}s^{v}_{N}(x,x^{\prime}))\vert u_{k}(y)\vert \vert u_{k}(y^{\prime})\vert
+A_{2}\int_{\vert {\bf k}\vert>\Lambda} d{\bf k}\exp(-\omega_{k}s^{v}_{N}(x,x^{\prime}))
\end{array}\end{equation}
where in the integral over ${\bf k}$ we set $\omega_{k}=\sqrt{{\bf k}^{2}+m^{2}}$.
The integral over ${\bf k}$ is decreasing exponentially as a function of $s_{N}$.

We make the estimate precise in the simple model of ${\cal N}=R^{D-d}$ (eq.(25)).
We  apply Weyl theory \cite{taylor} saying
that for large eigenvalues the sum over eigenvalues of the
Laplace-Beltrami operator on a d-dimensional compact manifold ${\cal
M}$ can be approximated by a d-dimensional integral with
$\epsilon_{k}\simeq {\bf k}^{2}$. If additionally we assume $\vert
u_{k}\vert\leq C$ then in eq.(25)
\begin{equation}\begin{array}{l} \vert({\cal
G}^{m}-{\cal G}^{m}_{0})(X,X^{\prime})\vert \leq
A_{1}\sum_{k<\Lambda}
 \vert c_{k}(x-x^{\prime})\vert\vert u_{k}(y)\vert \vert u_{k}(y^{\prime})\vert  + R_{\Lambda}(x-x^{\prime})
\end{array}\end{equation}
where from the Weyl approximation
\begin{equation}\begin{array}{l}
R_{\Lambda}(x-x^{\prime})=C^{2}\int_{0}^{\infty}d\tau\int_{\vert
{\bf k}\vert>\Lambda} d{\bf k}
 \exp(-\frac{1}{2}({\bf k}^{2}+m^{2})\tau
-\frac{1}{2\tau}(x-x^{\prime})^{2})(2\pi\tau)^{-\frac{D-d}{2}} \cr
=A\int_{0}^{\infty}d\tau \tau^{-\frac{d}{2}}\Gamma
(\frac{d}{2},\frac{1}{2}\tau\Lambda^{2})
 \exp(-\frac{1}{2}m^{2}\tau
-\frac{1}{2\tau}(x-x^{\prime})^{2})(2\pi\tau)^{-\frac{D-d}{2}}
\end{array}\end{equation}
where $\Gamma$ denotes the incomplete gamma function \cite{grad}. From the asymptotic
expansion of $\Gamma$ we obtain
 that
\begin{equation}
 R_{\Lambda}(x-x^{\prime})\simeq \exp(-\sqrt{\Lambda^{2}+m^{2}}\vert x-x^{\prime}\vert)
 \end{equation}
 for large $\vert x-x^{\prime}\vert$.
Hence, we can conclude that
 ${\cal G}^{m}-{\cal G}^{m}_{0}$ for large $\vert x-x^{\prime}\vert$
 is decreasing as $\exp(-\sqrt{m^{2}+\epsilon}\vert x-x^{\prime}\vert)$, where
 $\epsilon$ is the lowest non-zero eigenvalue of $-\triangle_{M}$.
The decay of $\vert({\cal
G}^{m}-{\cal G}^{m}_{0})(X,X^{\prime})\vert$ for large distances is determined by the first term on the rhs of
eq.(46).

 We  repeat the estimates in the model (25)
 without any reference to the Weyl approximation
  in the simplest case of ${\cal M}=S^{\beta}$ where $S^{\beta}$ is the
circle of radius   $\beta $ ( then the spectrum of $-\triangle_{M}$
is known). The method of an explicit sum over eigenvalues is simple
if $m=0$. Then, from eqs.(9) and (17)
\begin{equation}\begin{array}{l} {\cal G}^{0}(X,
X^{\prime})\cr =(2\pi\beta)^{-1}\int_{0}^{\infty}d\tau
(2\pi\tau)^{-\frac{D-1}{2}}\exp(-\frac{1}{2\tau}(x-x^{\prime})^{2})
\sum_{k}\exp(-\frac{\tau}{2}(\frac{  k}{\beta})^{2})\exp(i\frac{
k}{\beta}(y-y^{\prime})) \cr =\int_{0}^{\infty}d\tau
(2\pi\tau)^{-\frac{D}{2}}\exp(-\frac{1}{2\tau}(x-x^{\prime})^{2})
\sum_{k}\exp(-\frac{1}{2\tau}(y-y^{\prime}- 2\pi\beta k)^{2})
\end{array}\end{equation}Performing the integral
over $\tau$ we obtain
\begin{equation}\begin{array}{l}
{\cal G}^{0}(X
,X^{\prime})=(2\pi)^{-D+2}\Gamma(\frac{D}{2}-1)\sum_{k}((x-x^{\prime})^{2}+
(y-y^{\prime}- 2\pi\beta k)^{2})^{-\frac{D}{2}+1}
\end{array}\end{equation}
In order to perform the sum we apply the formula
\begin{equation}\begin{array}{l} \sum_{k}(\sigma^{2}+
(y-y^{\prime}-2\pi\beta k)^{2})^{-1}\cr =\frac{1}{4\beta\sigma}
\Big( \coth(\frac{1}{2\beta}(\sigma+i(y-y^{\prime}))+
\coth(\frac{1}{2\beta}(\sigma-i(y-y^{\prime}))\Big)\end{array}
\end{equation}
The formula (51) can be applied directly to eq.(50) with
$\sigma^{2}=(x-x^{\prime})^{2}$ if $D=4$. When $D$ is even and
bigger then $4$ then we differentiate eq.(51) over $\sigma^{2}$
and subsequently apply to the sum (50). If $D=2n+1$ is odd then we
have to use differentiation $n-1$ times together with an
integration in order to perform the sum in eq.(50).

We show this
procedure for $D=3$. This dimension is relevant for the model (25) of sec.3 and in
models of sec.4. We use the integral
\begin{equation} \int_{0}^{\infty} dr(r^{2}+a^{2})^{-1}=\pi a^{-1}
\end{equation}
in order to represent the Green function in $D=3$ dimensions in
the form\begin{equation}\begin{array}{l}{\cal G}^{0}(X,X^{\prime})\cr
=2^{-\frac{3}{2}}\pi^{-2}\beta^{-1}\int_{0}^{\infty}dr\frac{1}{4\sigma}
\Big( \coth(\frac{1}{2\beta}(\sigma+i(y-y^{\prime}))+
\coth(\frac{1}{2\beta}(\sigma-i(y-y^{\prime}))\Big)\end{array}
\end{equation}
where $\sigma^{2}=r^{2}+(x_{0}-x_{0}^{\prime
})^{2}+(x_{1}-x_{1}^{\prime })^{2}$. Then,
\begin{equation}\begin{array}{l}({\cal G}^{0}-{\cal G}^{0}_{0})(X,X^{\prime})\cr
=2^{-\frac{3}{2}}\pi^{-2}\beta^{-1}\int_{0}^{\infty}dr\frac{1}{4\sigma}
\Big( \coth(\frac{1}{2\beta}(\sigma+i(y-y^{\prime})))-1+
\coth(\frac{1}{2\beta}(\sigma-i(y-y^{\prime}))-1\Big)\end{array}
\end{equation}
where ${\cal G}^{0}_{0}= G^{0}$ is defined in eq.(26) with $D-d=2$ (
for $m=0$ and $D-d=2$ the integral (26) is divergent at large $\tau$;
this is the well-known  infrared problem for massless scalar fields,
it can be avoided by a choice of test functions for the smeared out
fields with no support at the zero momentum, in such a case the
integral (26) is defined as a logarithm of the distance ).

Applying the formula
\begin{displaymath}
\coth v-1=2\exp(-2v)(1-\exp(-2v))^{-1}
\end{displaymath}
we bound the integral on the rhs of eq.(54)
by\begin{displaymath}\vert({\cal G}^{0}-{\cal
G}^{0}_{0})(X,X^{\prime})\vert \leq
A\int_{0}^{a}dr\sigma^{-1}\exp(-2\sigma)=A
K_{0}\Big(\frac{1}{\beta}\sqrt{(x_{0}-x_{0}^{\prime
})^{2}+(x_{1}-x_{1}^{\prime })^{2}}\Big)
\end{displaymath}
for large $(x_{0}-x_{0}^{\prime
})^{2}+(x_{1}-x_{1}^{\prime })^{2}$,
where $K_{\nu}$ denotes the modified Bessel function of the third kind \cite{grad},
which is exponentially decreasing for large arguments.

 We conclude from eq.(54) that
the logarithmic two-dimensional propagator well approximates the
three dimensional propagator for distances $ (x_{0}-x_{0}^{\prime
})^{2}+(x_{1}-x_{1}^{\prime })^{2}>>\beta^{2}$. Applying
eqs.(50)-(51) we  could  confirm for any dimension $D-d$ the result
following from eqs.(46)-(48) that the difference ${\cal G}^{0}-{\cal
G}^{0}_{0}$ is exponentially small for large distances (the length
scale is $\beta$ which is equal to the radius of the circle
$S^{\beta}$ or in other words it is the square root of the inverse
of the lowest non-zero eigenvalue of $-\triangle_{M}$). For $D>3$
the Green function ${\cal G}$ would be approximated by ${\cal
G}^{0}_{0}\simeq \vert x-x^{\prime}\vert^{-D+3}$ ( by $\log \vert
x-x^{\prime}\vert $ in $D=3$). The high temperature limit of
interacting field theories is discussed in
\cite{jackiw}\cite{pisarski}.

\section{Quantum free fields on the product manifold}

 We introduce now a free Euclidean field as a
random field with the two-point correlation function equal to the
Green function (see \cite{jaffe}). The Green function (17) defines
the Gaussian Euclidean field
\begin{equation}
\Phi(x,y)\equiv\sum_{k}\Phi_{k}=\sum_{k}\chi_{k}(x)u_{k}(y)
\end{equation}
where
\begin{equation}
\langle
\chi_{k}(x)\chi_{r}(x^{\prime})\rangle=\delta_{kr}g_{k}^{m}(x,x^{\prime})
\end{equation}
$g_{k}^{m}$ as a Green function of the second order differential operator
is non-negative. This is a positive definite bilinear form. Hence, it
defines Gaussian  Euclidean field $\phi_{k}$ on ${\cal N}$.
In the example (25) of ${\cal N}=R^{D-d}$ we have $g_{k}^{m}=(-\triangle
+\omega_{k}^{2})^{-1}$.Then, $\phi_{k}^{m}$ is the Euclidean free field on $R^{D-d}$ with the mass
$\omega_{k}$ ($\phi_{0}^{m}$ has the mass $m$).
In an analytic continuation of the model (25) to the Minkowski space
$\chi_{k}$ becomes the free quantum field with the mass $\omega_{k}$ on the $D-d$ dimensional Minkowski space.

The models of sec.4
 have an expansion
\begin{equation}\begin{array}{l}
\Phi(x_{0},x_{1},y)\equiv\sum_{k}\Phi_{k}=\int
dp_{1}\exp(ip_{1}x_{1})\sum_{k,n}a_{k}(p_{1},n)
\phi_{n}^{k}(x_{0}) u_{k}(y)\cr
=\Phi_{0}(x_{0},x_{1})+\sum_{k>0}\Phi_{k}(x_{0},x_{1},y)\end{array}
\end{equation}
where
\begin{equation}
\langle
\overline{a}_{k}(p_{1},n)a_{k^{\prime}}(p_{1}^{\prime},n^{\prime})\rangle=
\delta(p_{1}-p_{1}^{\prime})\delta_{nn^{\prime}}\delta_{kk^{\prime}}
(p_{1}^{2}+\lambda_{n}(k)^{2})^{-1}\end{equation} We have
\begin{equation}\begin{array}{l}
\langle\Phi_{0}(x_{0},x_{1})\Phi_{0}(x_{0}^{\prime},x_{1}^{\prime})\rangle=
 {\cal G}_{0}^{m}(x,x^{\prime})
=\int_{0}^{\infty}d\tau\exp(-\frac{1}{2}m^{2}\tau
-\frac{1}{2\tau}(x-x^{\prime})^{2})(2\pi\tau)^{-1}
\end{array}\end{equation}
 and for $k>0$
\begin{equation}\begin{array}{l}
\langle\overline{\Phi}_{k}(x_{0},x_{1},y)\Phi_{k}(x_{0}^{\prime},x_{1}^{\prime},y^{\prime})\rangle\cr
=\frac{1}{\pi}\sum_{n}\lambda_{n}(k)^{-1}\exp(-\lambda_{n}(k)\vert
x_{1}-x_{1}^{\prime}\vert)\overline{\phi}_{n}^{k}(x_{0})\phi_{n}^{k}(x_{0}^{\prime})\overline{u}_{k}(y)u_{k}(y^{\prime})
\end{array}\end{equation}
We can continue analytically the Green functions (60) of the model (57)
into imaginary
values  of $x_{1}$ (then $x_{1}$ plays the role of time; the analytic continuation follows from the positivity
of the Green function (60) under
a reflection  of  $x_{1}$ \cite{jaffe})

\begin{equation}\begin{array}{l}
{\cal G}^{m}(x_{0},ix_{1},y,ix_{1}^{\prime},y^{\prime})\cr
=\frac{1}{\pi}\sum_{k,n} \lambda_{n}(k)^{-1}\exp(i\lambda_{n}(k)(
x_{1}-x_{1}^{\prime}))
\overline{\phi}_{n}^{k}(x_{0})\phi_{n}^{k}(x_{0}^{\prime})
\overline{u}_{k}(y)u_{k}(y^{\prime})\end{array}\end{equation} Such Green functions result
from the quantization in the Fock space
\begin{equation}\begin{array}{l}
\Phi(x_{0},x_{1},y)=\sum_{k,n}\exp(-i\lambda_{n}(k)x_{1})a_{k}(n)
\phi_{n}^{k}(x_{0}) u_{k}(y)\cr
+\sum_{k,n}\exp(i\lambda_{n}(k)x_{1})a_{k}^{+}(n)
\phi_{n}^{k}(x_{0}) \overline{ u}_{k}(y)\end{array}
\end{equation}
where
\begin{equation}
[a_{k}(n),a_{k^{\prime}}^{+}(n^{\prime})]=\frac{1}{\pi}
\delta_{nn^{\prime}}\delta_{kk^{\prime}}
\lambda_{n}(k)^{-1}\end{equation} are the creation and
annihilation operators.

In order to define a quantum field  from the Euclidean field (60) with another time
  we would need the reflection positivity in another direction .
  Various analytic continuations are possible on the same manifold ${\cal N}$. As an example,
  if ${\cal N}$ is the hyperbolic space then the Euclidean field  on ${\cal N}$
  can be analytically continued either to a quantum field on
  DeSitter space or to the one defined on the AntiDeSitter space \cite{jaffe}.
The model (38) of sec.4 ($a_{1}\simeq x_{0}^{2}$ ) with ${\cal
M}=R^{d}$ as the Euclidean version of the Rindler space has an
analytic continuation to the Rindler space or to the Milne space.
The model is also related by a conformal transformation to the
direct product of the real line
 and the hyperbolic space
(the models and their analytic continuations are discussed in \cite{hyper} and in the references cited there).

  In general, if we are able to define a quantum field with the Green function defined
by the analytic continuation of $g_{k}^{m}$ (18) (for every $k$) then we can define the
quantum field on an analytic continuation of ${\cal P}$ by means of the expansion (55).
\section{Discussion}
We have investigated some aspects of  quantum free fields on
product manifolds ${\cal P}={\cal N}\times {\cal M}$. Such studies
have a long history. It is an old argument that on a compact
manifold ${\cal M}$ the heavy modes (KK-modes) decouple from the
zero mode (defining the dimensional reduction). In this paper we
have studied the problem in the configuration space. An
approximation of quantum field theory by a lower dimensional one
is not trivial because the theories in various dimensions have
different short distance behaviour. The dimensional reduction of
the Kaluza-Klein type can make sense only above a certain length
scale determined by the size of the compact manifold
( or equivalently by the inverse of the square root of the lowest non-zero
eigenvalue of $-\triangle_{ M}$). In sec.5 we
have estimated the difference ${\cal G}-{\cal G}_{0}$, where
${\cal G}_{0}$ is the $D-d$ dimensional Green function.
We have shown in some models that there is no dimensional reduction for a
non-compact ${\cal M}$ \cite{haba}. Nevertheless, a study of Green functions
with non-compact manifolds ${\cal M}$ can lead to some unexpected
results. The conventional behaviour $\vert
x-x^{\prime}\vert^{-D+2}$ in $D$ dimensions of the Green functions
of quantum massless fields can fail on a manifold. The Green
functions of the  model of sec.4 can be more singular
than it would follow from the dimension-dependent index $D-2$. A
change of the short distance behaviour has been shown in the brane
model of Dvali et al \cite{dvali}. Their model can be formulated
in the framework of sec.2 with $D-d=1$. The equation for the Green
function reads
\begin{equation}
(-\partial_{0}^{2}+a^{2}(x_{0})(-\triangle +m^{2})){\cal G}^{m}=\delta
\end{equation}
The Fourier transform of ${\cal G}$ in $D-1$ variables satisfies
an equation for the quantum mechanical Green function with the
potential $V={\bf p}^{2}a^{2}(x_{0})$. We have studied solutions
of the equations for Green functions in \cite{hababrane} by means
of path integrals. The short distance behaviour depends on the
singularity of $a^{2}(x_{0})$ for $x_{0}\rightarrow 0$. In the
model of ref.\cite{dvali}
\begin{equation}
a^{2}(x_{0})=\alpha+\beta\delta(x_{0})
\end{equation}
It follows from \cite{dvali}\cite{hababrane} that ${\cal G}(0,y,0,y^{\prime})\simeq
\vert y-y^{\prime}\vert^{-D+3}$. This is the behaviour of the Green
function in $D-1$ dimensions. Such a behaviour of the Green
function at short distances may be called a dimensional reduction
in spite of the continuous spectrum of $\triangle$ on ${\cal M}$.

Let us still consider the models with a singular metric of sec.4 in view of the results
of \cite{hababrane}. The equation for the Green function
reads
\begin{equation}
(-\hat{\partial}_{0}^{2}-\partial_{1}^{2}+\vert\hat{x}_{0}\vert^{\frac{2\gamma}{1-\gamma}}(-\triangle
+m^{2})){\cal G}^{m}=\hat{\delta}
\end{equation}where $\hat{x}_{0}$ has been called $\eta$  in \cite{hababrane}.
Applying the methods and results of ref.\cite{hababrane} we obtain
\begin{equation}
{\cal G}(0,x_{1},0;0,x_{1}^{\prime},0)\simeq \vert
x_{1}-x_{1}^{\prime}\vert^{\frac{-D+2}{1-\gamma}}
\end{equation}and
\begin{equation}
{\cal G}(0,0,y;0,0,y^{\prime})\simeq \vert y-y^{\prime}\vert^{-D+2}
\end{equation}
for small distances. Hence, the Green functions in the $x_{1}$
direction are more singular (if $\gamma>0$ )than the ones in
$R^{D}$, whereas the singularity in the $y$ direction is the
same as the one in the $D$ dimensional free field theory on a flat space.

 In the model of
Dvali et al  \cite{dvali} the long distance behaviour of massless
fields is the same as the one in $D$ dimensions. The dimensional
reduction discussed in the present paper concerns the long
distance behaviour. The short distance behaviour can be different
from the canonical one (with the index $D-2$) only if the metric
is singular.

An analytic continuation of Green functions of the models of sec.4
with a Killing vector on ${\cal N}$ as discussed in sec.6 follows
from the general theory of a quantization of the symmetry
generated by the Killing vector formulated in \cite{jaffe}. In the
  system of  coordinates adapted to this Killing vector
  the analytic continuation concerns the coordinate  $x_{1}$. It
  gives the Hilbert space, the Hamiltonian and a unitary
  evolution. There can be more reflection symmetries on a
  manifold. The Riemannian manifolds of sec.4 are invariant
  under the reflection $x_{0}\rightarrow -x_{0}$.
However, a reflection positivity (with respect to $x_{0}$),
necessary for a definition of the quantum field (with $x_{0}$
as an imaginary time), remains unclear.
The reflection positivity may have a physical meaning.
 Tunnelling between
different topologies of space-time may require a reflection
invariant Riemannian manifold \cite{gibbons}. Models with a prior
to the Big Bang evolution \cite{vene}\cite{crunch} seem to require
a Euclidean version which is invariant under a time reflection.

\end{document}